\documentclass[aps,prl,twocolumn,showpacs,superscriptaddress,groupedaddress]{revtex4}  
\usepackage{graphicx}  
\usepackage{dcolumn}   
\usepackage{bm}        
\usepackage{amssymb}   
\usepackage{amsmath}
\usepackage{tikz}

\newcommand{\ti}{\text{i}}

\newcommand{\tc}{\text{c}}

\newcommand{\im}{\text{Im}}

\newcommand{\tr}{\,\text{tr}\,}

\newcommand{\sub}[1]{_{\text{#1}}}

\newcommand{\cev}[1]{\reflectbox{\ensuremath{\vec{\reflectbox{\ensuremath{#1}}}}}} 
\newcommand{\intd}[3]{\int_{#1}^{#2}\text{d}#3\,}
\newcommand{\intdd}[3]{\int_{#1}^{#2}\text{d}^3#3\,}

\newcommand{\usphere}{\tikz{\draw (0,0) circle (2pt); \fill (0,0) circle (0.5pt);}}

\newcommand{\ev}[1]{\langle #1 \rangle}
\newcommand{\sca}{^{(1)}}
\newcommand{\bu}{^{(0)}}

\newcommand{\rone}{\mathbf{r}_1}
\newcommand{\rtwo}{\mathbf{r}_2}

\DeclareMathAlphabet{\mathbfsf}{\encodingdefault}{\sfdefault}{bx}{n}

\begin{document}



\title{Correct zero-point energy-momentum of the electromagnetic field in media for studying the Casimir force}
\author{Friedrich Anton Burger}
\affiliation{Physikalisches Institut, Albert-Ludwigs-Universit\"at Freiburg, Hermann-Herder-Str. 3, 79104 Freiburg, Germany}
\author{Johannes Fiedler}
\affiliation{Physikalisches Institut, Albert-Ludwigs-Universit\"at Freiburg, Hermann-Herder-Str. 3, 79104 Freiburg, Germany}
\author{Stefan Yoshi Buhmann}
\affiliation{Physikalisches Institut, Albert-Ludwigs-Universit\"at Freiburg, Hermann-Herder-Str. 3, 79104 Freiburg, Germany}
\affiliation{Freiburg Institute for Advanced Studies, Albert-Ludwigs-Universit{\"a}t Freiburg, Albertstr. 19, 79104 Freiburg, Germany}

\date{\today}

\begin{abstract}
We address the long-standing controversy regarding the correct description of the electromagnetic energy-momentum tensor in media and its consequences for the Casimir force. The latter being due to the zero-point momentum of the electromagnetic field, competing approaches based on the Abraham or Maxwell stress tensor lead to different predictions. We consider the test scenario of two colloidal spherical particles submerged in a dielectric medium and use three criteria to distinguish the two approaches: we show that the Abraham stress tensor, and not the Maxwell stress tensor, leads to a Casimir force that is form-equivalent to Casimir-Polder and van der Waals forces, obeys duality as a fundamental symmetry and is consistent with microscopic many-body calculations.
\end{abstract}

\maketitle
\begin{figure}
	\includegraphics[width=0.5\textwidth]{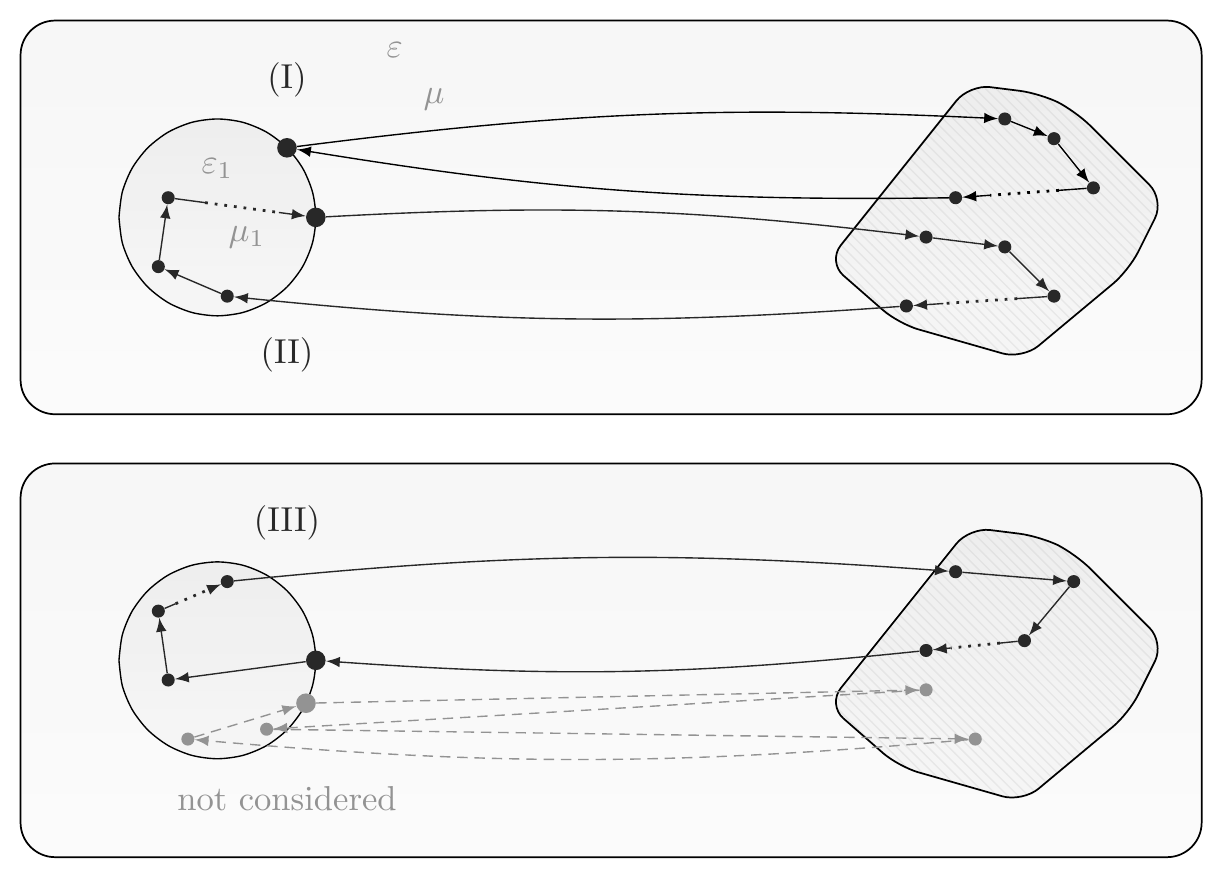}
	\caption{For a small sphere, all relevant paths contributing to the Green's tensor either only involve scattering events in the background bodies (I), or involve scattering events in the background bodies and subsequently in the small sphere (II), or the other way round (III).}
	\label{4}
\end{figure}
\begin{figure}
	\includegraphics[width=0.5\textwidth]{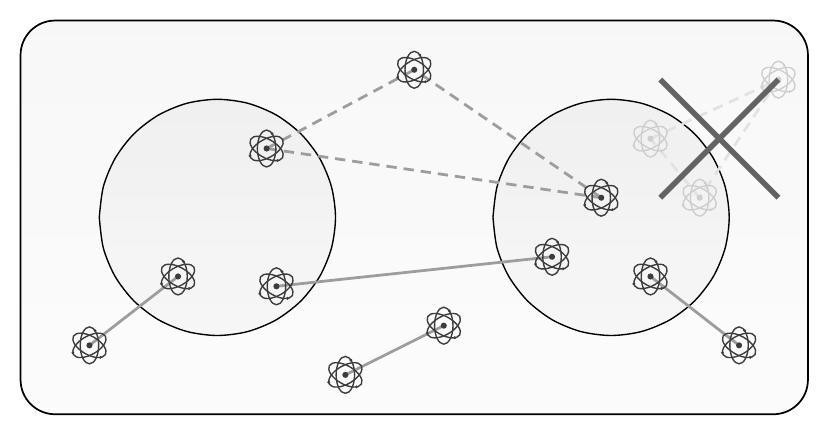}
	\caption{For comparing $C_{6,\text{A[M]}}$ with $C_{6,\text{Ham}}$ and $C_{6,\text{3p}}$, all terms of quadratic order in $\chi_1$, $\chi_2$ and $\chi$, representing all possible van der Waals interactions between molecules in the two spheres and the medium, are taken into account. In the third order, only the term proportional to $\chi_1\,\chi_2\,\chi$ is connected to $U_{3p}$, given by Eq.~\eqref{12}. Other contributions, such as the depicted term which is proportional to $\chi_2^2\,\chi$, are neglected for this comparison.}
	\label{11}
\end{figure}

The energy and momentum of the electromagnetic field in media have been subject to an old debate which has recently attracted renewed interest~\cite{hehl,pfeifer,richter,toptygin,rubinsztein}. This is due to the very fundamental significance of these quantities in the context of optical forces on particles in media, which prominently feature in experimental schemes for measuring Casimir forces~\cite{maianeto} or the critical Casimir effect~\cite{gambassi}, but are also relevant to the predicted Casimir force itself~\cite{dzyaloshinskii,raabe} as well as a hypothesised vacuum Casimir momentum transfer~\cite{van_tigellen,donaire}. Early works amongst others go back to Abraham~\cite{abraham} and Minkowski~\cite{minkowski}. 
For the case of isotropic media (we  also restrict our analysis to isotropic media in this letter), Abraham proposed a stress tensor~\cite{brevik_1979}
\begin{align}
\begin{split}
	\mathbf{T}\sub{A}(\mathbf{r},t)=&\mathbf{D}(\mathbf{r},t)\otimes\mathbf{E}(\mathbf{r},t)+\mathbf{H}(\mathbf{r},t)\otimes\mathbf{B}(\mathbf{r},t)\\
	&-\frac{1}{2}\Big[\mathbf{D}(\mathbf{r},t)\cdot\mathbf{E}(\mathbf{r},t)+\mathbf{H}(\mathbf{r},t)\cdot\mathbf{B}(\mathbf{r},t)\Big]\mathbb{I},
\end{split}
\label{15}
\end{align}
describing the flux of field momentum, with Poynting vector $\mathbf{S}\sub{A}(\mathbf{r},t)=\mathbf{E}(\mathbf{r},t)\times\mathbf{H}(\mathbf{r},t)$ describing the flux of field energy, energy density $w\sub{A}(\mathbf{r},t)=-\tr \mathbf{T}\sub{A}(\mathbf{r},t)$ and field momentum density $\mathbf{g}\sub{A}(\mathbf{r},t)=\mathbf{S}\sub{A}(\mathbf{r},t)/\tc^2$. Minkowski supported the same results for the first three quantities, but proposed a field momentum $\mathbf{g}\sub{Min}(\mathbf{r},t)=\mathbf{D}(\mathbf{r},t)\times\mathbf{B}(\mathbf{r},t)$ instead of the Abraham field momentum. This is the subject of the Abraham--Minkowski controversy.\\
Experiments on the momentum density and the stress tensor have mostly been performed by measuring the deformation of dielectric fluids in the electromagnetic field, for example when exposed to a laser beam~\cite{casner,theory_casner}, or inside a capacitor~\cite{hakim,goetz,brevik_1979}. The electromagnetic force density on charges depends on the stress tensor as well as the momentum density according to 
\begin{equation}
\mathbf{f}(\mathbf{r},t)=\nabla\cdot\mathbf{T}(\mathbf{r},t)-\dot{\mathbf{g}}(\mathbf{r},t).
\label{1}
\end{equation}
However, following such methods, a direct measurement of the field momentum in media is usually not possible, as $\dot{\mathbf{g}}$ averages to zero for  a stationary experimental setup~\cite{astrath,brevik_momentum}. From a more fundamental point of view the Abraham vs Minkowski momentum has also been analysed in the context of light forces on single atoms~\cite{hinds}. \\
The question of the correct energy and momentum of the electromagnetic field in media is also important for the subject of dispersion forces where one calculates the mean forces between bodies in media in presence of the zero-point electromagnetic field.
The time derivative of the momentum density $\dot{\mathbf{g}}$ in Eq.~\eqref{1} does not contribute and the dispersion force on a body of volume $V$ thus reads
\begin{align}
	\mathbf{F}=\intdd{V}{}{r}\ev{\hat{\mathbf{f}}(\mathbf{r})}=\intd{\partial V}{}{\mathbf{A}}\cdot\ev{\hat{\mathbf{T}}(\mathbf{r})}.
	\label{2}
\end{align}
Dzyaloshinskii, Lifshitz and Pitaevskii first came up with a theory capable of describing dispersion forces between plates immersed in homogeneous media at mechanical equilibrium~\cite{dzyaloshinskii}, generalising the result by Lifshitz for plates in vacuum~\cite{lifshitz}. Mechanical equilibrium implies that forces on the medium itself are balanced by pressure gradients within the medium. Dzyaloshinskii and co-workers claimed that the Abraham--Minkowski stress tensor~\eqref{15} is the correct choice for determining the Casimir force~\eqref{2} under these assumptions (for simplicity, we will only refer to it as Abraham stress tensor in the following). \\
In 2005, Raabe and Welsch proposed an alternative way to derive dispersion forces between bodies in media~\cite{raabe}. Calculating the Lorentz force on the internal charges and currents of a body, they found the dispersion force between bodies in media to be given by Eq.~\eqref{2} together with the Maxwell stress tensor
\begin{align}
\begin{split}
\mathbf{T}\sub{M}(\mathbf{r},t)=&\varepsilon_0\mathbf{E}(\mathbf{r},t)\otimes\mathbf{E}(\mathbf{r},t)+\frac{1}{\mu_0}\mathbf{B}(\mathbf{r},t)\otimes\mathbf{B}(\mathbf{r},t) \\
&-\frac{1}{2}\Big[\varepsilon_0\mathbf{E}^2(\mathbf{r},t)+\frac{1}{\mu_0}\mathbf{B}^2(\mathbf{r},t)\Big]\mathbb{I},
\end{split}
\label{14}
\end{align}
traditionally regarded as the stress tensor for the free-space electromagnetic field~\cite{jackson}.
Their result is in clear contradiction to the one of Dzyaloshinskii, Lifshitz and Pitaevskii. Note that works by Philbin based on canonical quantisation~\cite{philbin} and by Brevik and Ellingsen based on classical electrodynamics~\cite{brevik_comment} favour the Abraham stress tensor.
Raabe and Welsch regard their theory to be in consensus with considerations from classical electromagnetism~\cite{hehl} and argue that their approach, being purely electromagnetic, is on the same footing as microscopic dispersion forces such as van der Waals and Casimir--Polder forces~\cite{raabe06,raabe_comment,raabe_commentx2}.\\
Pitaevskii has argued that the approach by Raabe and Welsch is not complete due to its disregarding of the medium pressure and is thus not able to describe equilibrated media~\cite{pitaevskii_comment}. In contrast, Raabe and Welsch view the latter point as an advantage of their theory, as their not pressure-corrected dispersion forces may constitute a framework for analysing Casimir-induced pressures within the intervening medium. In a dilute-gas limit, such forces would emerge as the well-known Casimir--Polder forces between the medium atoms and the bod- \\ies. \\
Casimir forces between bodies in media have also been measured, using a surface force apparatus~\cite{horn, afshar, claesson} or atomic force microscope~\cite{ducker,senden,capasso}, often with a special focus on repulsive forces~\cite{hutter, milling, capasso_nat}. However, due to uncertainties in the force measurements and the optical data required to theoretically predict the forces, it is unclear whether current experiments are able to unambiguously discriminate between the two models.\\
In this article, we instead present a theoretical analysis where we investigate the question of the correct approach in a pragmatic way. We analyse the dispersion interaction between two small magneto-dielectric spheres in a homogeneous magneto-dielectric medium for the two theories and check which of the two approaches is on the same footing as the microscopic dispersion forces. Small spheres are well studied~\cite{hamaker,langbein,sambale}, play an important role in science, e.g. in colloid physics~\cite{boinovich}, and are particularly well suited for comparing macroscopic to microscopic dispersion forces: small spheres are only dipole-polarisable and are thus directly comparable to microscopic dipole-polarisable objects. Furthermore, the assumption of smallness facilitates the calculations and leads to more transparent results. \\
We employ three criteria: \textit{(i)} the correspondence to the Casimir--Polder and van der Waals force, \textit{(ii)} the behaviour under duality transformation and \textit{(iii)} the consistency with the microscopic theory based on summing van der Waals and Axilrod--Teller interactions. Finally, we also analyse whether the approach by Raabe and Welsch leads to forces on a medium in presence of a single sphere. This could indicate that their force, which unlike the force by Dzyaloshinskii, Lifshitz and Pitaevskii does not include balancing forces, is not buoyancy-corrected (the notion of buoyancy can be understood here in analogy to the case of a body inside water in the gravitational field). \\
Employing the quantisation scheme introduced in~\cite{buh1}, one finds for the zero-point expectation values of the electric-field stress tensor components
\begin{align}
\begin{split}
	\ev{\mathbf{E}(\mathbf{r})\otimes\mathbf{E}(\mathbf{r})}=&\frac{\hbar}{\varepsilon_0\pi}\intd{0}{\infty}{\omega}\frac{\omega^2}{\tc^2}\im\mathbf{G}(\mathbf{r},\mathbf{r}',\omega)\Big\vert_{r'=r} \\
	\ev{\mathbf{D}(\mathbf{r})\otimes\mathbf{E}(\mathbf{r})}=&\frac{\hbar}{\pi}\intd{0}{\infty}{\omega}\frac{\omega^2}{\tc^2}\im\left\{\varepsilon(\mathbf{r},\omega)\mathbf{G}(\mathbf{r},\mathbf{r}',\omega)\right\}\Big\vert_{r'=r} \\
\end{split}
\end{align}
and analogous results for the magnetic fields. $\mathbf{G}(\mathbf{r},\mathbf{r}',\omega)$ denotes the electromagnetic Green's tensor which describes the propagation of the electromagnetic field from a source point $\mathbf{r}'$ to a field point $\mathbf{r}$. It is the solution to the differential equation
\begin{equation}
	\left[\nabla\times\frac{1}{\mu(\mathbf{r},\omega)}\nabla\times-\frac{\omega^2}{\tc^2}\varepsilon(\mathbf{r},\omega)\right]\mathbf{G}(\mathbf{r},\mathbf{r}',\omega)=\boldsymbol{\delta}(\mathbf{r}-\mathbf{r}').
\end{equation}
 We subtract the bulk part $\mathbf{G}\bu$ of the Green's tensor, which is connected to the position-independent Lamb shift~\cite{lamb} and does not contribute to forces between bodies, and only use the scattering part $\mathbf{G}\sca$ in the following: $\mathbf{G}\to\mathbf{G}-\mathbf{G}\bu=\mathbf{G}\sca$. Also performing a Wick rotation on the complex frequency plane~\cite{wick},  Eq.~\eqref{2} with the Abraham and the Maxwell stress tensors becomes
\begin{align}
\begin{split}
	\mathbf{F}\sub{[A]M}=&-\frac{\hbar}{\pi}\intd{0}{\infty}{\xi}\intd{}{}{\mathbf{A}}\cdot\Bigg\{\frac{\xi^2}{\tc^2}[\varepsilon(\mathbf{r},\ti\xi)]\mathbf{G}\sca(\mathbf{r},\mathbf{r},\ti\xi)\\
	&+\frac{1}{[\mu(\mathbf{r},\ti\xi)]}\nabla\times\mathbf{G}\sca(\mathbf{r},\mathbf{r}',\ti\xi)\times\cev{\nabla}\Big\vert_{r'=r}\\
	&-\frac{1}{2}\mathbb{I}\tr\Bigg(\frac{\xi^2}{\tc^2}[\varepsilon(\mathbf{r},\ti\xi)]\mathbf{G}\sca(\mathbf{r},\mathbf{r},\ti\xi)\\
	&++\frac{1}{[\mu(\mathbf{r},\ti\xi)]}\nabla\times\mathbf{G}\sca(\mathbf{r},\mathbf{r}',\ti\xi)\times\cev{\nabla}\Big\vert_{r'=r}\Bigg)\Bigg\},
\end{split}
	\label{8}
\end{align}
respectively. The results for the two stress tensors only differ via the factors of $\varepsilon$ and $1/\mu$ in square brackets which only occur when using the Abraham stress tensor. \\
To calculate the dispersion force between two small spheres in a homogeneous medium, one has to find the scattering Green's tensor $\mathbf{G}\sca(\mathbf{R},\mathbf{R},\ti\xi)$ for this geometry, where $\mathbf{R}$ denotes positions on the surface of the sphere we are integrating over. To this end, we use the method of Born expansion to obtain the Green's tensor of a small sphere with the position arguments on its surface in presence of other background bodies and the medium. Within this method, one has to sum over all possible propagations of (virtual) photons from a source point to a field point to obtain the Green's tensor~\cite{buhborn}. The Born expansion for an electric body (volume $V$, permittivity $\varepsilon\sub{B}$) in a homogeneous dielectric ($\varepsilon$) reads (for brevity, the frequency arguments are omitted in the following)
\begin{align}
\begin{split}
	&\mathbf{G}\sca(\mathbf{r},\mathbf{r}')=\sum\limits_{k=1}^\infty(-1)^k\frac{\xi^{2k}}{\tc^{2k}}\left(\Delta \varepsilon\right)^{k}\!\!\intdd{V}{}{s_1}\!\dots\!\intdd{V}{}{s_k}\\
	&\times\mathbf{G}\bu(\mathbf{r},\mathbf{s}_1)\cdot\mathbf{G}\bu(\mathbf{s}_1,\mathbf{s}_2)\dots\mathbf{G}\bu(\mathbf{s}_k,\mathbf{r}'),
\end{split}
\end{align}
with $\Delta\varepsilon=\varepsilon\sub{B}-\varepsilon$. An analogous formula can be given for magnetic bodies. As the sphere is small compared to the distance to other bodies, the contribution from paths involving multiple scattering between the bodies and the small sphere are negligible. Three different classes of photon paths remain. A representative of each class and also an example of a negligible path are depicted in Fig.~\ref{4} for one background body.
Summing all paths from the three classes, one obtains the Green's tensor
\begin{align}
\begin{split}
	&\mathbf{G}_{\usphere_1}\sca(\mathbf{R},\mathbf{R})=\mathbf{G}\sca(\mathbf{R},\mathbf{R}) \\
	&-\frac{\xi^2\alpha^*_1}{\tc^2\varepsilon_0}\Big[\mathbf{G}\bu(\mathbf{R},\rone)\cdot\mathbf{G}\sca(\rone,\mathbf{R})\\
	&\;\;+\mathbf{G}\sca(\mathbf{R},\rone)\cdot\mathbf{G}\bu(\rone,\mathbf{R})\Big]\\
	&-\mu_0\beta^*_1\Big[\mathbf{G}\bu(\mathbf{R},\rone)\times\cev{\nabla}_{1}\cdot\nabla_{1}\times\mathbf{G}\sca(\rone,\mathbf{R})\\
	&\;\;+\mathbf{G}\sca(\mathbf{R},\rone)\times\cev{\nabla}_{1}\cdot\nabla_{1}\times\mathbf{G}\bu(\rone,\mathbf{R})\Big].
\end{split}
\label{5}
\end{align}
The first term, the Green's tensor for the background bodies, arises from class (I), the  term in the second line arises from class (II) and the term in the third line arises from class (III). The remaining terms come from the analogue magnetic contribution. $\rone$ denotes the centre of the sphere. $\alpha^*_1$ and $\beta^*_1$ are the electric and magnetic excess dipole polarisabilities of the sphere~\cite{mclachlan}, given by
\begin{equation}
	\alpha^*_1=4\pi\varepsilon_0\varepsilon R_1^3\frac{\varepsilon_1-\varepsilon}{\varepsilon_1+2\varepsilon}\text{ ~and~ }	\beta^*_1=\frac{4\pi R_1^3}{\mu_0\mu}\frac{\mu_1-\mu}{\mu_1+2\mu}.
	\label{10}
\end{equation}
Here, $\varepsilon_1$ and $\mu_1$ denote the sphere's permittivity and permeability, while $\varepsilon$ and $\mu$ are the respective properties of the medium. A second small sphere with centre at $\mathbf{r}_2$ is now introduced by specifying the background Green's tensor in Eq.~\eqref{5} to be the one of a second small sphere using the result of Sambale {\it et al.}~\cite{sambale}
\begin{align}
\begin{split}
	\mathbf{G}\sca_{\usphere_2}(\mathbf{r},\mathbf{r}')=&-\frac{\alpha^*_2\xi^2}{\varepsilon_0\tc^2}\mathbf{G}\bu(\mathbf{r},\rtwo)\cdot\mathbf{G}\bu(\rtwo,\mathbf{r}')\\
	&-\mu_0\beta^*_2\mathbf{G}\bu(\mathbf{r},\rtwo)\times\cev{\nabla}_2\cdot\nabla_2\times\mathbf{G}\bu(\rtwo,\mathbf{r}').
\end{split}
\label{6}
\end{align}
Inserting Eq.~\eqref{5} into Eq.~\eqref{8}, evaluating the surface integrals using leading order Taylor expansions in $\mathbf{R}-\mathbf{r}_1$,  specifying the background Green's tensor to be given by ~\eqref{6}, and by explicitly inserting the bulk Green's tensor in the non-retarded limit~\cite{buh1}, one obtains the dispersion forces between two small spheres in the non-retarded limit, where the spheres are close compared to the wavelengths of relevant transition frequencies. The corresponding dispersion potentials read
\begin{align}
\begin{split}
	U\sub{A[M]}(r_{12})=&\frac{C_{6,\text{A[M]}}}{r_{12}^6}\text{ with }\\
	C_{6,\text{A[M]}}=&\frac{-3\hbar}{16\pi^3}\intd{0}{\infty}{\xi}\Bigg(\frac{\alpha_1^*\alpha_2^*}{\varepsilon_0^2\varepsilon^{2\,[+1]}}+\mu_0^2\mu^{2[+1]}\beta^*_1\beta_2^*\Bigg)
\end{split}
\label{9}
\end{align}
and $r_{12}\equiv||\mathbf{r}_1-\mathbf{r}_2||$. The results for the two stress tensors only differ in the exponents of the medium response functions $\varepsilon$ and $\mu$, the additional powers in the square brackets only occur when using the Maxwell stress tensor. \\
Having found the force between two small spheres in a homogeneous medium on the basis of both stress tensors, we can now analyse the results by employing three criteria. \\
\textit{(i) Correspondence criterion:} The only properties of the two spheres entering Eq.~\eqref{9} are the excess dipole polarisabilities. By replacing these excess dipole polarisabilities by molecular ones, one should obtain the van der Waals potential of two molecules in a homogeneous medium. Explicitly inserting the bulk Green's tensor of a homogeneous medium in the non-retarded limit in Ref.~\cite{buh_vdw}, one sees that this is the case for the Abraham potential, but not for the Maxwell potential. \\
One may wonder whether it is possible to redefine the excess polarisabilities~(\ref{10}) such a way that the Maxwell potential obeys the correspondence criterion. However, this will lead to inconsistencies when requiring a formal correspondence to also hold for the Casimir--Polder potential. This can be seen by analysing the force on a small sphere in a medium in front of a perfectly reflecting surface, obtained by specifying the background Green's tensor in (\ref{5}) to represent this system. The resulting $C_3$ coefficients read
\begin{equation}
	C_{3,\text{A[M]}}=\frac{-\hbar}{16\pi^2}\intd{0}{\infty}{\xi}\left(\frac{\alpha_1^*}{\varepsilon_0\varepsilon^{1[+1]}}-\mu_0\mu^{1[+1]}\beta_1^*\right);
	\label{13}
\end{equation}    
the Abraham result again fits the corresponding Casimir--Polder result. It is not possible to rescale the excess polarisabilities~\eqref{10} in such a way that the Maxwell approach holds for Eqs.~\eqref{9} and~\eqref{13} simultaneously,   the former would demand the scaling factor $1/\sqrt{\varepsilon}$ for the electric excess polarisability whereas the latter would demand the factor $1/\varepsilon$.\\
\textit{(ii) Duality criterion:} The macroscopic Maxwell equations are symmetric with respect to a duality transformation which exchanges electric and magnetic fields~\cite{buh1}. This symmetry is also obeyed by the van der Waals force and the Casimir--Polder force: the electric and magnetic contributions transform into each other and the total forces are invariant. A macroscopic dispersion potential which also incorporates this symmetry is thus favourable. Exchanging the response functions $\varepsilon$ and $\mu$ for the medium and the two spheres in Eq.~\eqref{9} together with Eq.~\eqref{10}, one sees that the Abraham potential obeys duality invariance, while the Maxwell potential does not.  \\
\textit{(iii) Microscopic criterion:} Finally, we investigate which of the two potentials is consistent with the microscopic theory obtained by summing all microscopic potentials which are involved in the interaction between the two spheres. This analysis is restricted to the electric parts of the potentials. We take into account the two-particle van der Waals potentials and the three-particle Axilrod--Teller potentials. \\
The electric van der Waals potential between two molecules with polarisabilities $\alpha$ and $\alpha'$ is given by
\begin{equation*}
	U\sub{vdW}(\mathbf{r},\mathbf{r}')=\frac{-3\hbar}{16\pi^3\varepsilon_0^2}\intd{0}{\infty}{\xi}\alpha\alpha'\frac{1}{||\mathbf{r}-\mathbf{r}'||^6}.
\end{equation*}
The contribution of the van der Waals potentials to the potential between two spheres in a medium can be obtained by identifying the part of the energy arising from all van der Waals interactions which depends on the finite distance of the spheres. This method was found by Hamaker~\cite{hamaker}. It yields $U\sub{Ham}(r_{12})=C_{6,\text{Ham}}/r_{12}^6$ with
\begin{align}
\begin{split}
	C_{6,\text{Ham}}=&-\frac{\hbar R_1^3R_2^3}{3\pi}\intd{0}{\infty}{\xi}\Big[\eta_1\alpha_1\eta_2\alpha_2\\
	&-\eta_1\alpha_1\eta\alpha-\eta\alpha\eta_2\alpha_2+\eta^2\alpha^2\Big],
\end{split}
\end{align}
where $\alpha_1$, $\alpha_2$ and $\alpha$ are the polarisabilities of the molecules in the two spheres and the medium and $\eta_1$, $\eta_2$ and $\eta$ the corresponding molecular number densities. \\
The Axilrod--Teller potential between three molecules with polarisabilities $\alpha$, $\alpha'$ and $\alpha''$ is given by~\cite{at}
\begin{align*}
	U\sub{AT}(\mathbf{r},\mathbf{r}',\mathbf{r}'')=&\frac{3\hbar}{64\pi^4\varepsilon_0^3}\intd{0}{\infty}{\xi}\alpha\alpha'\alpha''\\
	&\frac{1+3\cos\theta\cos\theta'\cos\theta''}{||\mathbf{r}-\mathbf{r}'||^2||\mathbf{r}'-\mathbf{r}''||^3||\mathbf{r}''-\mathbf{r}||^3},
\end{align*}
where the angles are the ones of the triangle formed by the three molecules. The calculation of the three-particle contribution to the potential between two spheres in a medium for the case when two molecules are in separate spheres and the third one is in the medium yields $U\sub{3p}(r_{12})=C_{6,\text{3p}}/r_{12}^6$ with
\begin{align}
\begin{split}
C_{6,\text{3p}}=\frac{2\hbar R_1^3R_2^3}{9\pi\varepsilon_0^3}\intd{0}{\infty}{\xi}\eta_1\alpha_1\eta_2\alpha_2\eta\alpha.
\label{12}
\end{split}
\end{align}
The dispersion potential between two spheres, given by Eq.~\eqref{9} for the two approaches, should coincide with the sum of all microscopic potentials between the involved molecules. This identification is obtained by means of the Clausius--Mossotti law~\cite{clausius}, which links the macroscopic susceptibility $\chi$ to the molecular polarisability $\alpha$. For dilute bodies, it reads $\eta\alpha/\varepsilon_0=\chi$. Hence, an expansion of the $C_6$ coefficient in Eq.~\eqref{9} in orders of $\chi_1$, $\chi_2$ and $\chi$ should yield the microscopic contributions in this case. Up to the third order in $\chi_1$, $\chi_2$ and $\chi$ and by only taking into account the term proportional to $\chi_1\,\chi_2\,\chi$ in the third order as depicted in Fig.~\ref{11}, one obtains
\begin{align}
	C_{6,\text{A[M]}}=C_{6,\text{Ham}}+[5/2]C_{6,\text{3p}}.
\end{align}
Both, the Abraham and the Maxwell potential, yield the correct two-particle Hamaker contribution, while only the Abraham potential also comes with the correct three-particle contribution whereas the Maxwell potential overestimates it by a facter of $5/2$.\\
It could still be possible that the Maxwell potential between two small spheres in medium is not representing the full potential but excluding buoyancy corrections coming from the surrounding medium. However, this is not the case: the Maxwell potential, like the Abraham potential, vanishes when replacing one of the spheres by a spherical volume of medium since the corresponding excess polarisabilities in~\eqref{9} vanish in both cases.\\
The preceding analysis leads us to the conclusion that the approach by Dzyaloshinskii, Lifshitz and Pitaevskii on the basis of the Abraham stress tensor is confirmed for the geometry of two small spheres in a homogeneous medium. The force is of the same structure as the microscopic van der Waals and Casimir--Polder forces and is symmetric under exchange of electric and magnetic contributions. Furthermore, we have shown that it can at least up to three-particle interactions be viewed as the sum of all microscopic dispersion forces between the contributing molecules. \\
Our identification of the correct description of Casimir forces in media has an immediate impact on studies of the melting of bodies~\cite{melting} and the wetting of surfaces~\cite{wetting}. Here, one is interested in the dispersion interaction of a body, coated by a layer of melted material or liquid, with the surrounding air. Another recently considered application is the formation of ice crystals under water~\cite{ice}.  \\ 
The approach by Raabe and Welsch on the basis of the Maxwell stress tensor on the other hand is not fulfilling these criteria. Furthermore, contrary to their findings for a planar geometry, their approach does not lead to forces on an initially homogeneous medium in presence of a small sphere. By definition, this is also the case for the theory based on the Abraham stress tensor and it is thus not possible to calculate such forces on the medium using the present theory. The prediction of many-body dispersion forces in a medium as induced by a nearby solid body remains an important open question which could be relevant, inter alia, to the atmosphere capture by small asteroids. \\
We thank R.~Corkery, C.~Henkel and V.~Shatokhin for stimulating discussions. S.Y.B. and J.F. gratefully acknowledge support by the German Research Council (grant BU1803/3-1), the University of Freiburg (Research Innovation Fund) and the Freiburg Institute for Advanced Studies (S.Y.B.).

\end{document}